# Current distribution and giant magnetoimpedance in composite wires with helical magnetic anisotropy

N.A. Buznikov [a,b,*], A.S. Antonov [a], A.B. Granovsky [c], C.G. Kim [b], C.O. Kim [b], X.P. Li [d], S.S. Yoon [b,e]

[a] *Institute for Theoretical and Applied Electrodynamics, Russian Academy of Sciences, Moscow 125412, Russia*

[b] *Research Center for Advanced Magnetic Materials, Chungnam National University, Daejeon 305-764, Republic of Korea*

[c] *Faculty of Physics, M.V. Lomonosov Moscow State University, Moscow 119992, Russia*

[d] *Department of Mechanical Engineering and Division of Bioengineering, National University of Singapore, Singapore 119260, Singapore*

[e] *Department of Physics, Andong National University, Andong 760-749, Republic of Korea*

**Abstract**

The giant magnetoimpedance effect in composite wires consisting of a non-magnetic inner core and soft magnetic shell is studied theoretically. It is assumed that the magnetic shell has a helical anisotropy. The current and field distributions in the composite wire are found by means of a simultaneous solution of Maxwell equations and the Landau–Lifshitz equation. The expressions for the diagonal and off-diagonal impedance are obtained for low and high frequencies. The dependences of the impedance on the anisotropy axis angle and the shell thickness are analyzed. Maximum field sensitivity is shown to correspond to the case of the circular anisotropy in the magnetic shell. It is demonstrated that the optimum shell thickness to obtain maximum impedance ratio is equal to the effective skin depth in the magnetic material.

---

[*] Corresponding author. *E-mail address:* n_buznikov@mail.ru

# 1. Introduction

The giant magnetoimpedance (GMI) effect implies a huge change in the impedance of a magnetic conductor with the variation of an external magnetic field. This effect is promising for technological applications, in particular, for the development of sensors of weak magnetic fields and magnetic-field controlled devices. The GMI has been observed first in Co-based amorphous wires and ribbons [1–4]. It has been shown that the origin of the GMI can be understood in terms of the classical skin effect in a magnetic conductor [3], as a consequence of the change in the penetration depth of the AC current by the application of an external magnetic field.

More recently, the GMI has been observed in composite wires [5–16]. These wires consist of a highly conductive non-magnetic inner core (typically copper) and a soft magnetic shell and are fabricated by means of the electroplating, electrodeposition or cold-drawn technique. Experimental studies and theoretical estimations [17,18] have demonstrated that the presence of the highly conductive inner core in the composite wires results in a significant increase of the GMI effect. Moreover, the composite wires have some other advantages in comparison with amorphous wires, in particular, an arbitrary choice of the materials for the shell and core, a relatively simple fabrication technology and a possibility of use a high-amplitude AC current. It should be noted also that the non-magnetic inner core excludes the appearance of the complicated domain structures and domain walls between the surface and inner part of the wire. In this connection, the composite wires may have, in principle, more soft magnetic properties, which is key factor for magnetic-field sensors. In contrast to amorphous wires, the GMI effect in the composite wires is not restricted by the change of the skin depth under the influence of the external magnetic field [19]. In this case, the GMI effect is mainly driven by the inductance of the shell and the resistance of the core.

To understand the GMI in the composite wires it is fundamental to know the current distribution over the cross section of the sample. This is a key object in the problem of the determination of the optimum magnetic shell thickness to obtain high GMI response. The current distribution in the composite wires has been calculated as a function of the frequency by means of the finite element method [20]. In these calculations, it has been assumed that the permeability may be presented as a scalar value, and the effect of the external field has been simulated by gradually reducing of the sample permeability. Thus, this approach allows one to explain the current distribution and the field dependence of the impedance only qualitatively. The analytical solutions for the field distribution and the impedance in the composite wires have been obtained in Refs. [17,18] for the composite wires having the longitudinal anisotropy in the magnetic shell. These models explain quite well some experimental results on the GMI effect in the composite wires [18]. However, experimental data show that depending on the fabrication process the composite wires may have also the radial [9−11,14] and circular (or helical) magnetic anisotropy [4−6,12,15,16].

For the samples with circular (or helical) magnetic anisotropy it is essential to take into account the tensor form of the permeability. In this case, the field-dependent signal can be measured either as the voltage across the sample, or as the voltage in the pick-up coil wound around the sample [21,22]. The pick-up coil voltage response is related to the cross-magnetization process, which appears since the current induces an AC axial magnetization variation. This effect has been referred to as the off-diagonal magnetoimpedance. The off-diagonal magnetoimpedance has been studied in detail for Co-based amorphous wires [21−26], and it has been demonstrated that the off-diagonal impedance may be preferable for sensors applications since it generates a linear voltage response with enhanced field sensitivity. Recently, the off-diagonal magnetoimpedance has been observed also in composite wires [27].

The aim of the present paper is to develop the model to calculate the GMI effect in the composite wires with the helical magnetic anisotropy. It is assumed that the outer shell has no domain structure, and the model is based on the concept of the field and frequency dependent surface impedance tensor [23−26,28,29]. The distribution of the electric and magnetic fields within the composite wire is found as a function of the frequency and external magnetic field. Expressions for the diagonal and off-diagonal impedance components are obtained for the cases of low and high frequencies. The field and frequency dependences of the impedance are analyzed as a function of the anisotropy axis angle and shell thickness.

## 2. Basic equations

Let us consider a composite wire of diameter $D$ consisting of a highly conductive inner core of diameter $d$ and an outer soft magnetic shell. The AC current $I=I_0\exp(-i\omega t)$ flows along the wire (along $z$-axis), and the external DC magnetic field $H_e$ is parallel to the current. The calculation of the impedance is based on a solution of Maxwell equations for the electric and magnetic fields and the Landau–Lifshitz equation of motion for the magnetization. An analytical treatment is possible in a linear approximation with respect to the time-variable parameters and under assumption of a local relationship between the magnetic field and the magnetization. Further, we neglect a domain structure in the magnetic shell and assume that the permeability is determined by the magnetization rotation.

The distribution of the easy axes in the magnetic shell depends on the stresses induced in the wire during the fabrication process and may vary over the shell volume. We assume that the shell has the helical uniaxial anisotropy, and the anisotropy axis makes the angle $\psi$ with the transverse direction. To calculate the permeability we neglect the demagnetizing fields and the contribution of the exchange energy. In this case, the permeability tensor $\hat{\mu}$ can be found by means of the well-known procedure of the solution of the linearized Landau–Lifshitz equation and is given by [25]

$$\hat{\mu} = \begin{pmatrix} 1+\mu_1 & -i\mu_a \sin\theta & i\mu_a \cos\theta \\ i\mu_a \sin\theta & 1+\mu_2 \sin^2\theta & -\mu_2 \sin\theta \cos\theta \\ -i\mu_a \cos\theta & -\mu_2 \sin\theta \cos\theta & 1+\mu_2 \cos^2\theta \end{pmatrix}, \quad (1)$$

where

$$\mu_1 = \frac{\omega_m(\omega_1 - i\kappa\omega)}{(\omega_1 - i\kappa\omega)(\omega_2 - i\kappa\omega) - \omega^2},$$

$$\mu_2 = \frac{\omega_m(\omega_2 - i\kappa\omega)}{(\omega_1 - i\kappa\omega)(\omega_2 - i\kappa\omega) - \omega^2}, \quad (2)$$

$$\mu_a = \frac{\omega_m \omega}{(\omega_1 - i\kappa\omega)(\omega_2 - i\kappa\omega) - \omega^2},$$

$$\omega_m = \gamma 4\pi M,$$
$$\omega_1 = \gamma[H_a \cos^2(\theta-\psi) + H_e \sin\theta], \quad (3)$$
$$\omega_2 = \gamma[H_a \cos\{2(\theta-\psi)\} + H_e \sin\theta].$$

Here $M$ is the saturation magnetization, $H_a$ is the anisotropy field, $\gamma$ is the gyromagnetic constant, $\kappa$ is the Gilbert damping parameter and $\theta$ is the equilibrium angle between the magnetization vector and the transverse direction. The angle $\theta$ can be found by minimizing the free energy, which can be presented as a sum of the anisotropy energy and the Zeeman energy. The minimization procedure results in the equation for the equilibrium angle

$$H_a \sin(\theta - \psi)\cos(\theta - \psi) - H_e \cos\theta = 0. \tag{4}$$

Assuming that the electric and magnetic fields depend only on the radial coordinate and taking into account the cylindrical symmetry, we may write Maxwell equations within the non-magnetic core, $\rho < d/2$, in the following form:

$$\frac{\partial e_z^{(1)}}{\partial \rho} = -\frac{i\omega}{c} h_\varphi^{(1)}, \quad \frac{\partial e_\varphi^{(1)}}{\partial \rho} + \frac{e_\varphi^{(1)}}{\rho} = \frac{i\omega}{c} h_z^{(1)},$$

$$\frac{\partial h_z^{(1)}}{\partial \rho} = -\frac{4\pi\sigma_1}{c} e_\varphi^{(1)}, \quad \frac{\partial h_\varphi^{(1)}}{\partial \rho} + \frac{h_\varphi^{(1)}}{\rho} = \frac{4\pi\sigma_1}{c} e_z^{(1)}. \tag{5}$$

Here $\sigma_1$ is the core conductivity, $c$ is the velocity of light and the subscripts $\varphi$ and $z$ correspond to the circular and longitudinal components of the fields, respectively.

Within the magnetic shell, $d/2 < \rho < D/2$, Maxwell equations can be reduced to two coupled differential equations for the magnetic field components [22–25]

$$\frac{\partial^2 h_\varphi^{(2)}}{\partial \rho^2} + \frac{1}{\rho}\frac{\partial h_\varphi^{(2)}}{\partial \rho} - \frac{h_\varphi^{(2)}}{\rho^2} + \frac{2i}{\delta_2^2} \times \left[\mu_{\varphi\varphi} h_\varphi^{(2)} + \mu_{\varphi z} h_z^{(2)}\right] = 0,$$

$$\frac{\partial^2 h_z^{(2)}}{\partial \rho^2} + \frac{1}{\rho}\frac{\partial h_z^{(2)}}{\partial \rho} + \frac{2i}{\delta_2^2} \times \left[\mu_{\varphi z} h_\varphi^{(2)} + \mu_{zz} h_z^{(2)}\right] = 0. \tag{6}$$

Here $\delta_2 = c/(2\pi\sigma_2\omega)^{1/2}$, $\sigma_2$ is the conductivity of the magnetic shell, $\sigma_2 < \sigma_1$, and

$$\mu_{\varphi\varphi} = 1 + \mu_{ef} \sin^2\theta,$$
$$\mu_{zz} = 1 + \mu_{ef} \cos^2\theta, \tag{7}$$
$$\mu_{\varphi z} = -\mu_{ef} \sin\theta \cos\theta.$$

It follows from Eq. (7) that the magnetic properties of the composite wire are determined by the effective permeability $\mu_{ef}$ only. The parameter $\mu_{ef}$ is given by the relation [22,23,25]

$$\mu_{ef} = \frac{\omega_m(\omega_m + \omega_1 - i\kappa\omega)}{(\omega_m + \omega_1 - i\kappa\omega)(\omega_2 - i\kappa\omega) - \omega^2}. \tag{8}$$

The electric field in the magnetic shell can be found by means of the following equations:

$$e_\varphi^{(2)} = -\frac{c}{4\pi\sigma_2}\frac{\partial h_z^{(2)}}{\partial \rho},$$
$$e_z^{(2)} = \frac{c}{4\pi\sigma_2}\left[\frac{\partial h_\varphi^{(2)}}{\partial \rho} + \frac{h_\varphi^{(2)}}{\rho}\right]. \qquad (9)$$

The components of the electric and magnetic fields should satisfy the continuity conditions at the core–shell interface:

$$e_z^{(1)}(d/2) = e_z^{(2)}(d/2),$$
$$h_\varphi^{(1)}(d/2) = h_\varphi^{(2)}(d/2),$$
$$e_\varphi^{(1)}(d/2) = e_\varphi^{(2)}(d/2), \qquad (10)$$
$$h_z^{(1)}(d/2) = h_z^{(2)}(d/2).$$

Furthermore, the components of the magnetic field at the wire surface are determined by the excitation conditions and can be expressed as

$$h_\varphi^{(2)}(D/2) = 4I/cD,$$
$$h_z^{(2)}(D/2) = 0. \qquad (11)$$

Thus, the distribution of the electric and magnetic fields within the composite wire is described by Eqs. (5), (6), (9)–(11). The diagonal component of the wire impedance $Z_{zz}$ can be expressed through the diagonal component $\zeta_{zz}$ of the surface impedance tensor:

$$Z_{zz} = (4l/cD)\zeta_{zz} = (4l/cD) \times \left.\frac{e_z^{(2)}}{h_\varphi^{(2)}}\right|_{\rho=D/2}, \qquad (12)$$

where $l$ is the wire length.

Due to the existence of the circular electric field, the voltage in the pick-up coil wound around the wire has a non-zero value. The off-diagonal impedance $Z_{\varphi z}$ is defined as the ratio of the pick-up voltage to the current and is proportional to the off-diagonal component $\zeta_{\varphi z}$ of the surface impedance tensor [22–25]:

$$Z_{\varphi z} = (4\pi N/c)\zeta_{\varphi z} = (4\pi N/c) \times \left.\frac{e_\varphi^{(2)}}{h_\varphi^{(2)}}\right|_{\rho=D/2}, \qquad (13)$$

where $N$ is the number of turns in the pick-up coil.

The set of coupled Eqs. (6) for the magnetic field in the shell cannot be solved analytically in general case. Below, we find the asymptotic solutions for the fields distribution and the impedance in two limiting cases of high and low frequencies, when the effective skin depth in the magnetic shell is much less and much higher than the wire radius, respectively. Such approach allows one to describe adequately the field and frequency dependences of the impedance in the composite wires.

## 3. Wire impedance: high-frequency approximation

The general solution of the Maxwell equations (5) within the core has the form

$$e_z^{(1)}(\rho) = A J_0(k\rho),$$
$$h_\varphi^{(1)}(\rho) = (4\pi\sigma_1/ck) A J_1(k\rho),$$
$$e_\varphi^{(1)}(\rho) = B J_1(k\rho),$$
$$h_z^{(1)}(\rho) = (4\pi\sigma_1/ck) B J_0(k\rho),$$
(14)

where $J_0$ and $J_1$ are the Bessel functions of the first kind, $k=(1+\mathrm{i})/\delta_1$, $\delta_1 = c/(2\pi\sigma_1\omega)^{1/2}$, $A$ and $B$ are the constants.

In the case of high enough frequency, the effective skin depth in the magnetic shell is much less than the wire radius. Then, we may present the magnetic field in the shell in the following form [22–25]:

$$h_\varphi^{(2)}(\rho), h_z^{(2)}(\rho) \propto \exp\{\lambda(\rho - D/2)\}.$$
(15)

Taking into account only the leading terms with respect to $\lambda$ in Eqs. (6), we may obtain the values of the parameter $\lambda$ [23,25]:

$$\lambda_{1,2} = \pm(1-\mathrm{i})/\delta_2, \quad \lambda_{3,4} = \pm(1-\mathrm{i})\sqrt{\mu_{\mathrm{ef}}+1}/\delta_2.$$
(16)

The ratio between the amplitudes of the field components $h_\varphi^{(2)}$ and $h_z^{(2)}$ can be found from Eqs. (6), which results in

$$h_\varphi^{(2)}(\rho) = \cos\theta[C\exp\{\lambda_1(\rho-D/2)\} + D\exp\{\lambda_2(\rho-D/2)\}]$$
$$\quad + \sin\theta[E\exp\{\lambda_3(\rho-D/2)\} + F\exp\{\lambda_4(\rho-D/2)\}],$$
$$h_z^{(2)}(\rho) = \sin\theta[C\exp\{\lambda_1(\rho-D/2)\} + D\exp\{\lambda_2(\rho-D/2)\}]$$
$$\quad - \cos\theta[E\exp\{\lambda_3(\rho-D/2)\} + F\exp\{\lambda_4(\rho-D/2)\}].$$
(17)

Here $C$, $D$, $E$ and $F$ are the constants. The components of the electric field can be obtained by means of Eqs. (9) and (17):

$$e_z^{(2)}(\rho) = (c/4\pi\sigma_2)\cos\theta[C\lambda_1 \exp\{\lambda_1(\rho - D/2)\} + D\lambda_2 \exp\{\lambda_2(\rho - D/2)\}]$$
$$+ (c/4\pi\sigma_2)\sin\theta[E\lambda_3 \exp\{\lambda_3(\rho - D/2)\} + F\lambda_4 \exp\{\lambda_4(\rho - D/2)\}],$$
$$e_\varphi^{(2)}(\rho) = -(c/4\pi\sigma_2)\sin\theta[C\lambda_1 \exp\{\lambda_1(\rho - D/2)\} + D\lambda_2 \exp\{\lambda_2(\rho - D/2)\}]$$
$$+ (c/4\pi\sigma_2)\cos\theta[E\lambda_3 \exp\{\lambda_3(\rho - D/2)\} + F\lambda_4 \exp\{\lambda_4(\rho - D/2)\}]. \tag{18}$$

It follows from Eqs. (16)–(18) that the high-frequency approximation is valid if $\delta_2/(\mu_{\text{ef}}+1)^{1/2} \ll D/2$. It should be noted that in the case of the amorphous wire the solution for the electric and magnetic fields consists of two exponential terms only, with the positive sign of the real part of $\lambda$ [22,23,25]. On the contrary, for the composite wire we should take into account four different exponential terms, since the condition of the finite values of the fields at $\rho=0$ is not required.

The constants $A$, $B$, $C$, $D$, $E$ and $F$ can be found from the boundary conditions (10) and (11). The diagonal and off-diagonal impedance can be calculated as

$$Z_{zz} = \frac{l}{\pi\sigma_2 D} \times \frac{(C\lambda_1 + D\lambda_2)\cos\theta + (E\lambda_3 + F\lambda_4)\sin\theta}{(C+D)\cos\theta + (E+F)\sin\theta}, \tag{19}$$

$$Z_{\varphi z} = -\frac{N}{\sigma_2} \times \frac{(C\lambda_1 + D\lambda_2)\sin\theta - (E\lambda_3 + F\lambda_4)\cos\theta}{(C+D)\cos\theta + (E+F)\sin\theta}. \tag{20}$$

Note that in the limit of the strong skin-effect in the magnetic shell, $\delta_2/(\mu_{\text{ef}}+1)^{1/2} \ll (D-d)/2$, the expressions for the electric and magnetic fields can be simplified. It can be shown that in this case $A=B=D=F=0$, and the fields distribution is described by the well-known solution for the amorphous wire with helical anisotropy [23,25]. Correspondingly, Eqs. (19) and (20) for the diagonal and off-diagonal impedance have a more simple form [22–25,28–31]:

$$Z_{zz} = \frac{(1-i)l}{\pi\sigma_2 D\delta_2} \times \left[\sqrt{\mu_{\text{ef}}+1}\sin^2\theta + \cos^2\theta\right], \tag{21}$$

$$Z_{\varphi z} = \frac{(1-i)N}{\sigma_2 \delta_2} \times \left[\sqrt{\mu_{\text{ef}}+1}-1\right]\sin\theta\cos\theta. \tag{22}$$

## 4. Wire impedance: low-frequency approximation

In the case of low frequencies, the solution of Eqs. (6) can be found in the form of series [23,25]. It can be shown that the solution of Eqs. (6) can be presented as the arbitrarily linear combination of the series

$$h_\varphi^{(2)}(\rho) = CU_1 + DU_2 + EU_3 + FU_4,$$
$$h_z^{(2)}(\rho) = CV_1 + DV_2 + EV_3 + FV_4,$$
(23)

where functions $U_j$, $V_j$ ($j=1\div 4$) are determined by means of expressions

$$U_1 = \sum_n \alpha_n (2\rho/D)^n, \quad V_1 = \sum_n \beta_n (2\rho/D)^n, \quad n = 2k+1, \quad k = 0,1,\ldots,$$

$$U_2 = \sum_n \alpha_n (2\rho/D)^n, \quad V_2 = \sum_n \beta_n (2\rho/D)^n, \quad n = 2k, \quad k = 0,1,\ldots,$$

$$U_3 = D/2\rho + \sum_n (2\rho/D)^n \{a_n \log(2\rho/D) + c_n\}, \quad n = 2k+1, \quad k = 0,1,\ldots,$$

$$V_3 = \sum_n (2\rho/D)^n \{b_n \log(2\rho/D) + d_n\}, \quad n = 2k+1, \quad k = 0,1,\ldots,$$
(24)

$$U_4 = \sum_n (2\rho/D)^n \{a_n \log(2\rho/D) + c_n\}, \quad n = 2k, \quad k = 0,1,\ldots,$$

$$V_4 = \sum_n (2\rho/D)^n \{b_n \log(2\rho/D) + d_n\}, \quad n = 2k, \quad k = 0,1,\ldots.$$

It should be noted that in the case of $\mu_{yz}=0$, Eqs. (23) and (24) represent the Taylor expansion of the corresponding Bessel functions, which are the solutions of Eqs. (6) with $\mu_{yz}=0$. The coefficients in Eqs. (24) with $k>0$ can be calculated by means of recursion expressions. Using Eqs. (6), we obtain

$$\alpha_n = -\frac{iD^2}{2\delta_2^2(n^2-1)} \times (\mu_{\varphi\varphi}\alpha_{n-2} + \mu_{\varphi z}\beta_{n-2}),$$

$$\beta_n = -\frac{iD^2}{2\delta_2^2 n^2} \times (\mu_{\varphi z}\alpha_{n-2} + \mu_{zz}\beta_{n-2}),$$

$$a_n = -\frac{iD^2}{2\delta_2^2(n^2-1)} \times (\mu_{\varphi\varphi}a_{n-2} + \mu_{\varphi z}b_{n-2}),$$
(25)

$$b_n = -\frac{iD^2}{2\delta_2^2 n^2} \times (\mu_{\varphi z}a_{n-2} + \mu_{zz}b_{n-2}),$$

$$c_n = -\frac{iD^2}{2\delta_2^2(n^2-1)} \times (\mu_{\varphi\varphi}c_{n-2} + \mu_{\varphi z}d_{n-2}) - \frac{2na_n}{n^2-1},$$

$$d_n = -\frac{iD^2}{2\delta_2^2 n^2} \times (\mu_{\varphi z}c_{n-2} + \mu_{zz}d_{n-2}) - \frac{2b_n}{n}.$$

The initial coefficients can be found from the stationary solution of Eqs. (6) and are given by

$$\alpha_0 = 0, \quad \alpha_1 = 1, \quad \beta_0 = 1, \quad \beta_1 = 0,$$
$$a_0 = 0, \quad c_0 = 0, \quad b_0 = 1, \quad d_0 = 1, \tag{26}$$
$$a_1 = c_1 = -\mathrm{i}(D^2/\delta_2^2)\mu_{\varphi\varphi}, \quad b_1 = 0, \quad d_1 = 0.$$

Thus, at low frequencies, the magnetic field distribution in the wire can be found by means of Eqs. (14), (23)–(26). The constants $A$, $B$, $C$, $D$, $E$ and $F$ can be obtained from the boundary conditions (10) and (11). After that, the diagonal and off-diagonal components of the wire impedance can be calculated as

$$Z_{zz} = \frac{l}{\pi\sigma_2 D} \times \left.\frac{\frac{1}{\rho}\frac{\partial}{\partial\rho}[\rho(CU_1 + DU_2 + EU_3 + FU_4)]}{CU_1 + DU_2 + EU_3 + FU_4}\right|_{\rho=D/2}, \tag{27}$$

$$Z_{\varphi z} = -\frac{N}{\sigma_2} \times \left.\frac{\frac{\partial}{\partial\rho}[CV_1 + DV_2 + EV_3 + FV_4]}{CU_1 + DU_2 + EU_3 + FU_4}\right|_{\rho=D/2}. \tag{28}$$

## 5. Results and discussion

Thus, the distribution of the electric and magnetic fields within the composite wire can be calculated at any frequency by means of Eqs. (10), (11), (14), (17), (18), (23)–(26). As an example, Fig. 1 shows the distribution of the longitudinal electric field as a function of the radial coordinate at different frequencies $f=\omega/2\pi$. For the further calculations the parameters of the composite wire are taken as $M$=600 Gs, $H_a$=2 Oe, $\sigma_1$=5×10$^{17}$ s$^{-1}$, $\sigma_2$=10$^{16}$ s$^{-1}$ and $\kappa$=0.1. For the convenience, we reduce the electric field to the corresponding value of the field at the constant current, $e_0=(cDh_0/\pi)[\sigma_1 d^2 + \sigma_2(D^2-d^2)]^{-1}$. At low frequencies, the current flows mainly through the non-magnetic core. As the frequency increases, the electric field and, correspondingly, the current density in the core decrease gradually. At very high frequencies, the electric field in the core drops almost to zero, and the current flows mainly in the magnetic shell.

Fig. 2 illustrates the effect of the external magnetic field $H_e$ on the distribution of the electric field within the composite wire. With the growth of the external field, the permeability of the shell increases. This results in the increase of the electric field and current density in the shell. When the field exceeds some threshold value corresponding to the anisotropy field $H_a$, the permeability begins to decrease, and the electric field in the core increases. At high external fields, the current flows mainly in the non-magnetic core with higher conductivity.

The total current flowing through the core or magnetic shell can be calculated by means of the integration of the current density over the cross-section. Fig. 3 shows the real part of the total current flowing through the non-magnetic core $I_c$ as a function of the frequency at different values of the external magnetic field. It follows from Fig. 3 that the current in the core drops with the frequency increase, and the current redistribution between the core and the shell depends significantly on the external field, which is responsible for the GMI effect.

Shown in Fig. 4 are the field dependences of the diagonal and off-diagonal impedance for different angles of the anisotropy axis. The impedance values are reduced to the DC wire resistance, $R_{DC}=(4l/\pi)[\sigma_1 d^2+\sigma_2(D^2-d^2)]^{-1}$. Note that the results are presented only for the region of the positive fields, since the calculated curves are symmetrical with respect to the sign of the external field. It follows from Fig. 4 that at low values of the anisotropy axis angle $\psi$ the impedance increases with the field, achieves the maximum at $H_e \cong H_a$ and then decreases. If the anisotropy axis angle $\psi > \pi/4$, the impedance has the maximum at $H_e=0$ and decreases monotonically with the growth of $H_e$. It should be noted that the anisotropy axis angle may be changed by annealing of the composite wires, and the similar transition from the two-peak to single-peak field dependence of the impedance has been observed after annealing of the composite wires in the presence of the longitudinal magnetic field [16]. It is seen from Fig. 4 that the field sensitivity of both the diagonal and off-diagonal impedance increases with the decrease $\psi$, and the maximum sensitivity attains in the case of the circular anisotropy, $\psi=0$.

Fig. 5 shows the effect of the magnetic shell thickness on the field dependence of the impedance. At a fixed frequency, the impedance ratio increases with the shell thickness and reaches its maximum near some critical value. For a higher shell thickness, the amplitude of the impedance ratio becomes lower (see dotted lines in Fig. 5). The frequency dependence of the maximum magnetoimpedance ratios, $\Delta Z_{zz}$ and $\Delta Z_{\varphi z}$, is shown in Fig. 6 for different values of the shell thickness. These ratios are defined as the difference between the peak impedance value, $Z_{max}$, and the impedance at zero external field, $Z(0)$. It follows from Fig. 6 that at high frequencies the composite wire with thick magnetic shell has lower impedance ratio. The results obtained can be explained as follows. With the increase of the frequency, the skin depth in the magnetic shell decreases. At some critical frequency the skin depth becomes lower than the ferromagnetic layer thickness. In this case, the AC current flows almost through the shell, and the resistance of the composite wire is high enough.

The results of calculations show that the optimum shell thickness is equal to the effective skin depth in the magnetic material, $(D-d)_{opt}/2 = \delta_2/(\mu_{ef}+1)^{1/2}$. Note that this result coincides with that obtained previously by using simple estimations [17,19]. The optimum shell thickness is shown in Fig. 7 as a function of the frequency at different values of the anisotropy axis angle $\psi$. It follows from Fig. 7 that at sufficiently low frequencies the optimum shell thickness increases sharply with $\psi$. At high frequencies, the optimum shell thickness is sufficiently low and depends slightly on the anisotropy axis angle.

Thus, the model developed allows one to describe main features of the field and frequency dependences of both the diagonal and off-diagonal magnetoimpedance in the composite wires with the helical anisotropy. It is assumed above that the permeability is determined by the magnetization rotation only. This approximation is valid for sufficiently high frequencies, when the domain-walls motion is damped by the eddy currents. At low frequencies (of the order of 1 MHz and lower), the single-peak behavior of the field dependence of the diagonal impedance has been observed at any value of the anisotropy axis angle [16] due to the effect of the domain-walls motion. The contribution of the domain-walls motion to the effective permeability and impedance response at low frequencies can be found by the methods described in Refs. [32–34]. It should be noted also that in real composite wires the anisotropy field $H_a$ and the anisotropy axis angle $\psi$ may change over the magnetic shell thickness. To take into account this fact, the shell of the composite wire can be subdivided into several layers with slightly varying $H_a$ and $\psi$. This may result in a certain averaging of the magnetoimpedance response of the composite wire.

Note also that in this paper we restrict our consideration to the case of the excitation of the composite wire by the AC current. The approach proposed allows one to take into account also the effect of a weak longitudinal AC magnetic field on the impedance. In this case, all expressions for the electric and magnetic field remains the same, and the only difference is the change of the boundary condition (11) for the longitudinal magnetic field at the wire surface [24,25]. Moreover, in the framework of the present model the asymmetric impedance response due to the presence of the DC bias current may be calculated similar to the case of the Co-based amorphous wires [24–26,28,29]. To take into account the effect of the bias current the corresponding modification in Eq. (4) for the equilibrium magnetization angle should be done.

The results obtained above are in a qualitative agreement with experimental data on the GMI effect in the composite wires. However, the theory predicts higher field sensitivity in comparison with that obtained in experiments. Probably, this is related to the technological problems in the fabrication of the composite wires. The electroplating and the electrodeposition do not provide for soft magnetic properties due to the defects at the surface, whereas the cold-drawn technique does not allow obtaining the circular magnetic anisotropy.

In conclusion of this section, it should be noted that we assume above that the AC current amplitude is sufficiently low, and the voltage response is proportional to the wire impedance. At higher AC current amplitudes, the relation between the sample magnetization and the current amplitude becomes nonlinear. In this case, higher harmonics appear in the voltage response measured across the wire or in the pick-up coil [5,27,35–38]. Under nonlinear conditions, the field sensitivity of higher harmonics in the voltage response turns out to be much higher as compared to the sensitivity of the first harmonic, which can be very useful for magnetic-field sensors. The behavior of the voltage spectra at high current amplitudes is related to the magnetization reversal of the part of the wire and can be described within the framework of the quasi-static Stoner−Wohlfarth model [5,37].

**6. Conclusions**

In this paper, the model to calculate the field distribution and the magnetoimpedance in the composite wires with the helical magnetic anisotropy is presented. In the single-domain approximation, asymptotic solutions of Maxwell equation are found for low and high frequencies. Based on these solutions, the variation of the current and field distributions with the frequency and external magnetic field is analyzed. The analytical expressions for the magnetoimpedance are found in terms of the field and frequency dependent surface impedance tensor. It is shown that the maximum impedance ratio attains in the composite wires having the circular magnetic anisotropy. The optimum shell thickness is analyzed as a function of the frequency and anisotropy axis angle and is shown to be equal to the effective skin depth in the magnetic material. The results obtained may be useful for understanding the GMI effect in composite wires.

**Acknowledgements**

This work was supported by the Korea Science and Engineering Foundation through ReCAMM. N.A. Buznikov would like to acknowledge the support of the Brain Pool Program.

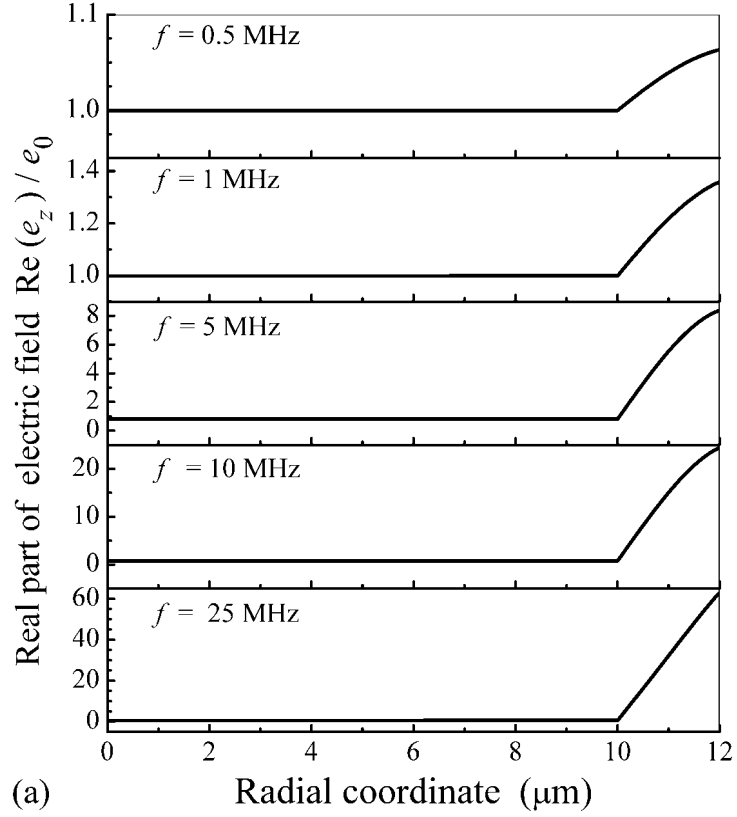

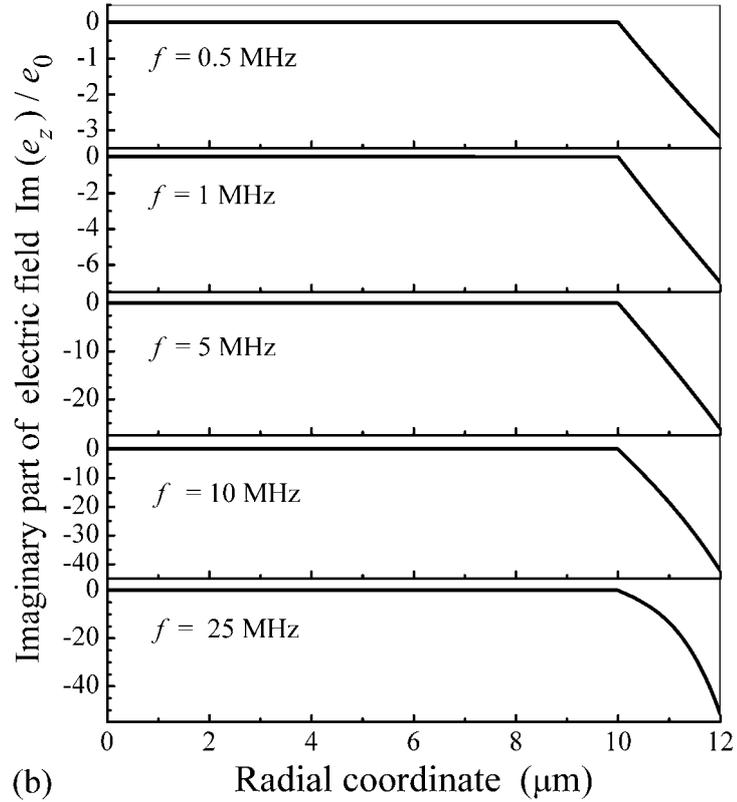

Fig. 1. Dependences of the real (a) and imaginary (b) parts of the longitudinal electric field on the radial coordinate at $H_e=1$ Oe and different frequencies. Parameters used for calculations are $D=24$ μm, $d=20$ μm, $M=600$ Gs, $H_a=2$ Oe, $\psi=0.1\pi$, $\sigma_1=5\times10^{17}$ s$^{-1}$, $\sigma_2=10^{16}$ s$^{-1}$, $\kappa=0.1$.

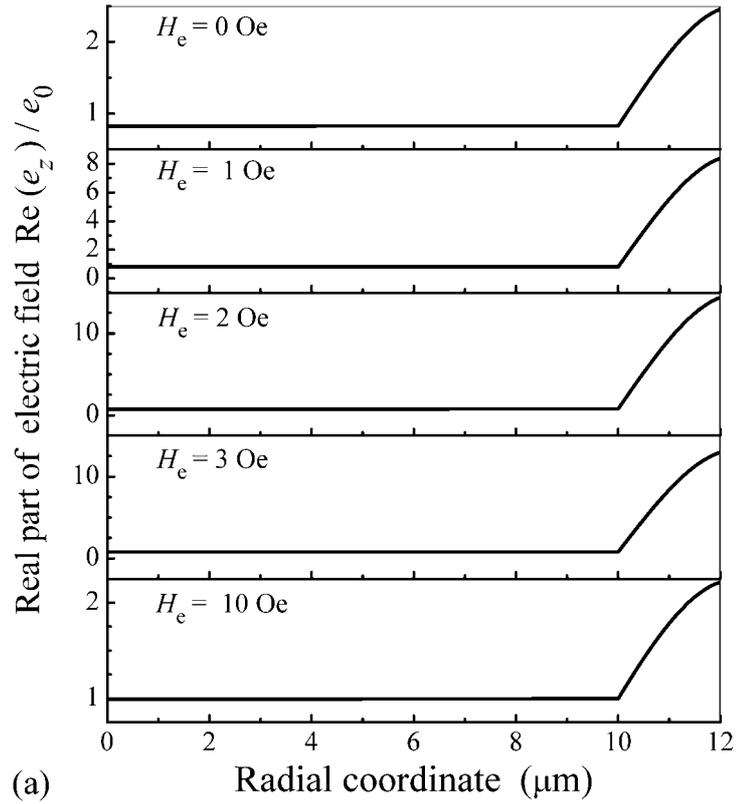

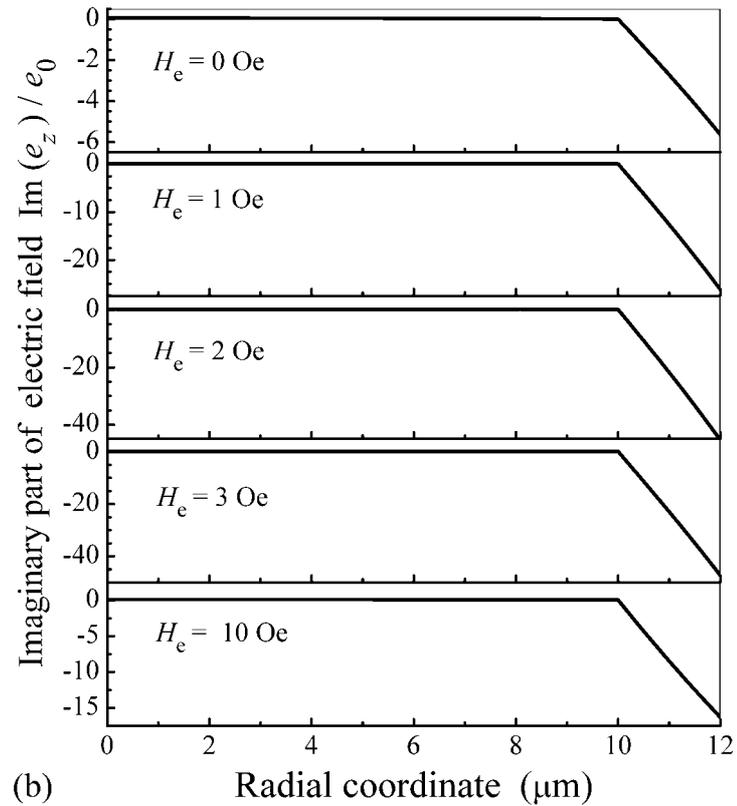

Fig. 2. Dependences of the real (a) and imaginary (b) parts of the longitudinal electric field on the radial coordinate at $f=5$ MHz and different external fields. Parameters are the same as in Fig. 1.

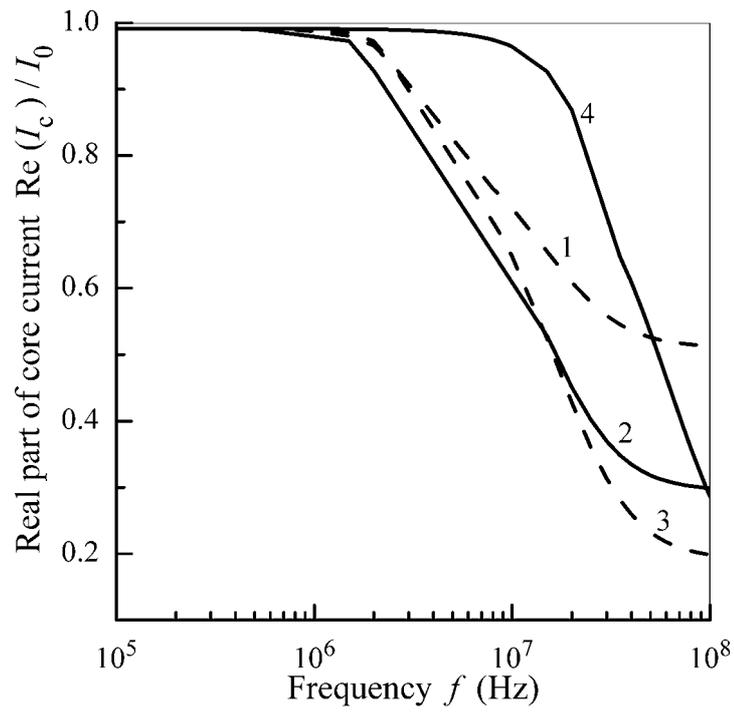

Fig. 3. Dependence of the real part of the current flowing through the core $I_c$ on the current frequency at different external field $H_e$: (1) $H_e=1\,\mathrm{Oe}$; (2) $H_e=2\,\mathrm{Oe}$; (3) $H_e=3\,\mathrm{Oe}$; (4) $H_e=10\,\mathrm{Oe}$. Parameters are the same as in Fig. 1.

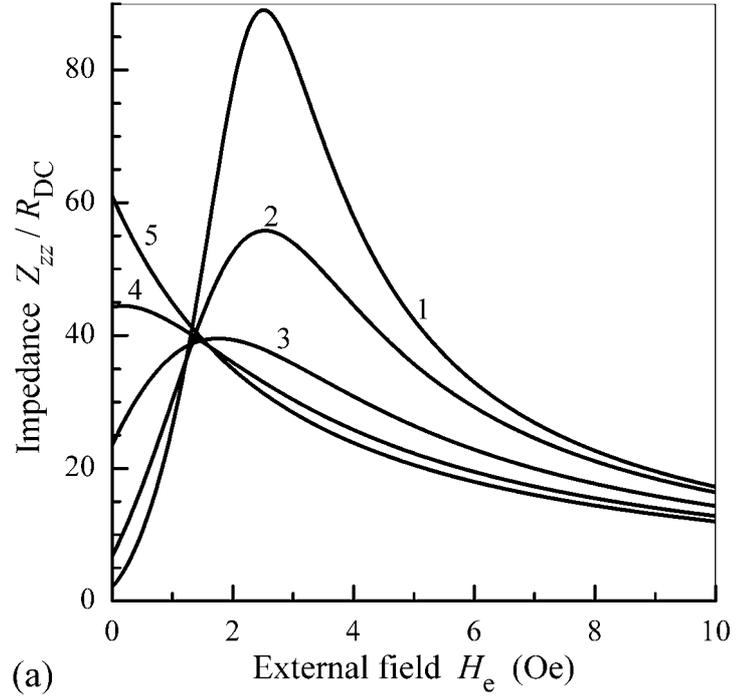

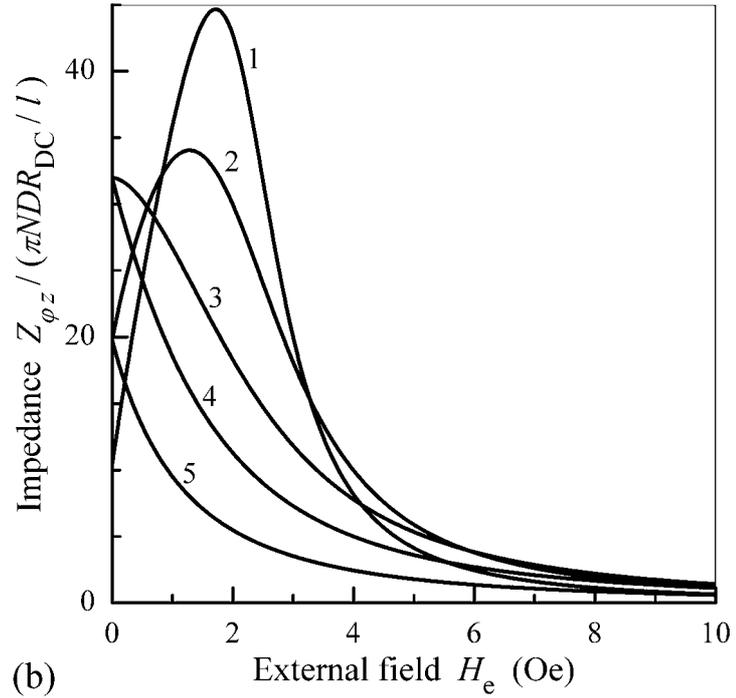

Fig. 4. Diagonal (a) and off-diagonal impedance (b) versus external field at $f=10$ MHz and different anisotropy axis angle $\psi$: (1) $\psi=0.05\pi$; (2) $\psi=0.1\pi$; (3) $\psi=0.2\pi$; (4) $\psi=0.3\pi$; (5) $\psi=0.4\pi$. Parameters used for calculations are $D=22\,\mu$m, $d=20\,\mu$m, $M=600$ Gs, $H_a=2$ Oe, $\sigma_1=5\times10^{17}\,\mathrm{s}^{-1}$, $\sigma_2=10^{16}\,\mathrm{s}^{-1}$, $\kappa=0.1$.

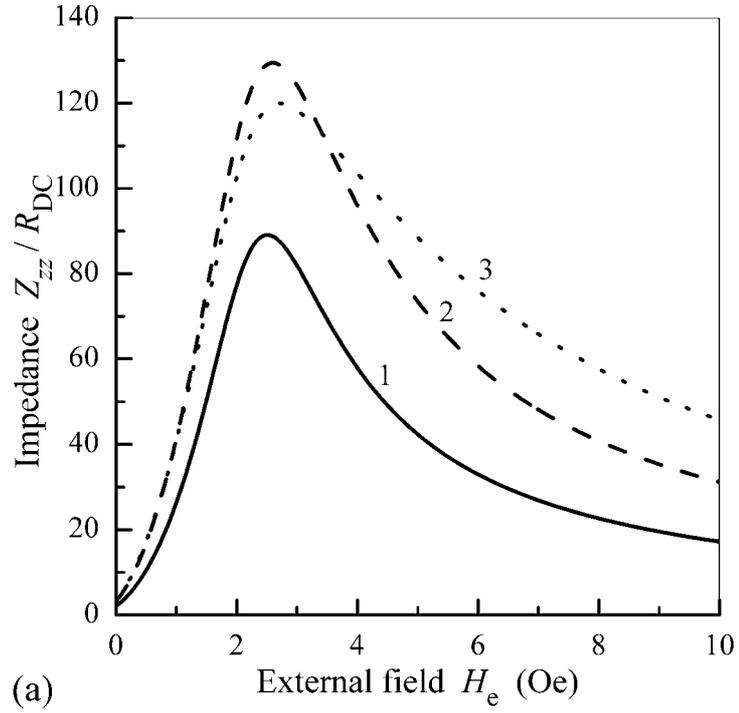

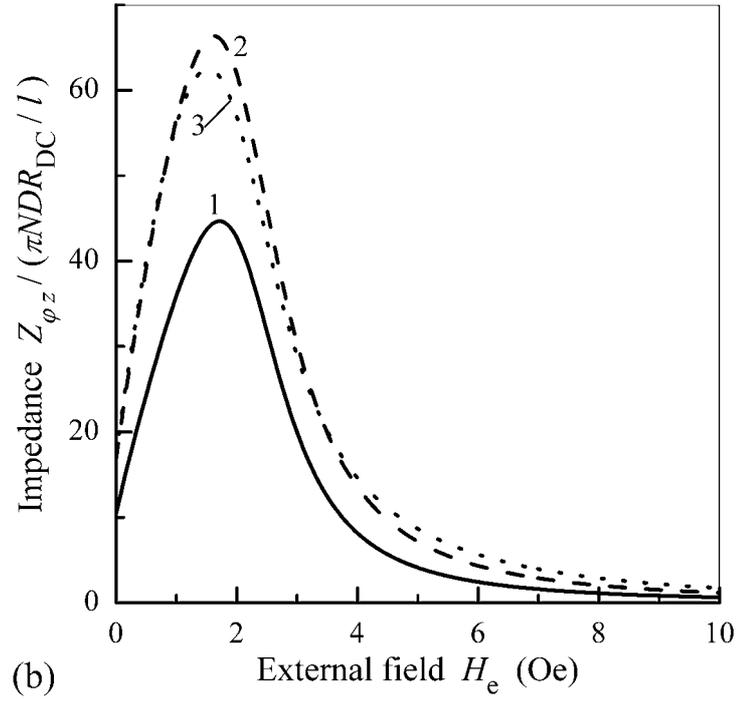

Fig. 5. Diagonal (a) and off-diagonal impedance (b) versus external field at $f=10$ MHz and different wire diameter $D$: (1) $D=22$ μm; (2) $D=24$ μm; (3) $D=27$ μm. Parameters used for calculations are $d=20$ μm, $M=600$ Gs, $H_a=2$ Oe, $\psi=0.05\pi$, $\sigma_1=5\times10^{17}$ s$^{-1}$, $\sigma_2=10^{16}$ s$^{-1}$, $\kappa=0.1$.

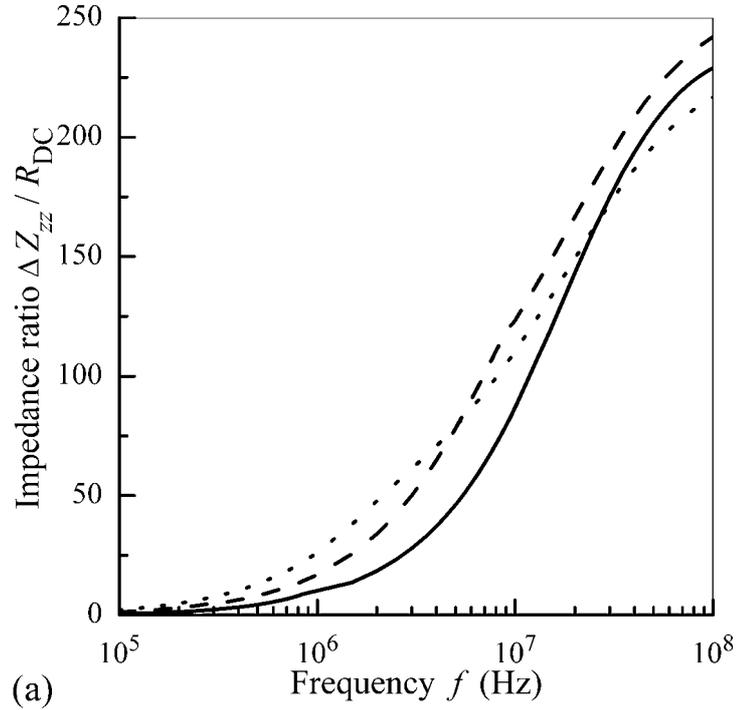

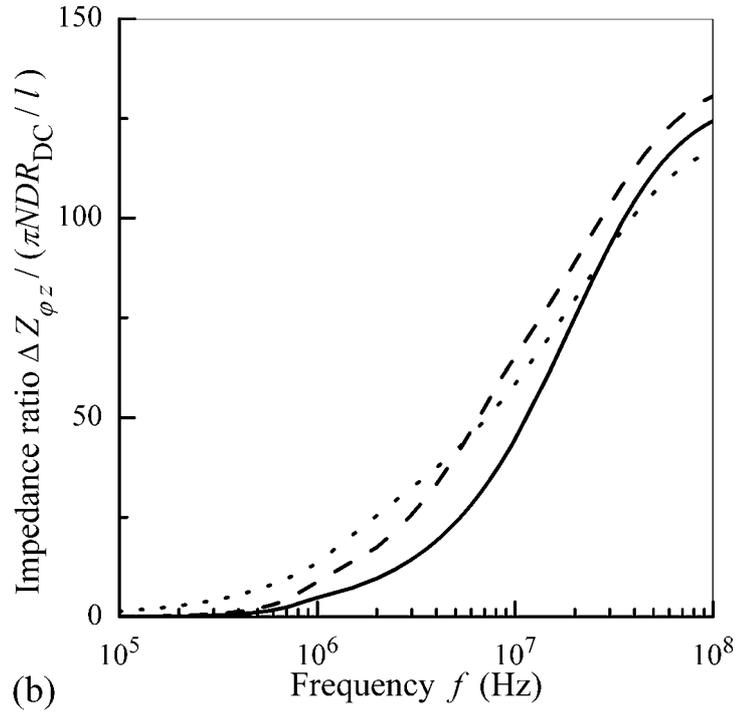

Fig. 6. Dependences of the impedance ratio for diagonal (a) and off-diagonal impedance (b) on the current frequency at different wire diameter $D$: solid lines, $D=22\,\mu m$; dashed lines, $D=24\,\mu m$; dotted lines, $D=27\,\mu m$. Parameters used for calculations are $d=20\,\mu m$, $M=600\,Gs$, $H_a=2\,Oe$, $\psi=0.05\pi$, $\sigma_1=5\times10^{17}\,s^{-1}$, $\sigma_2=10^{16}\,s^{-1}$, $\kappa=0.1$.

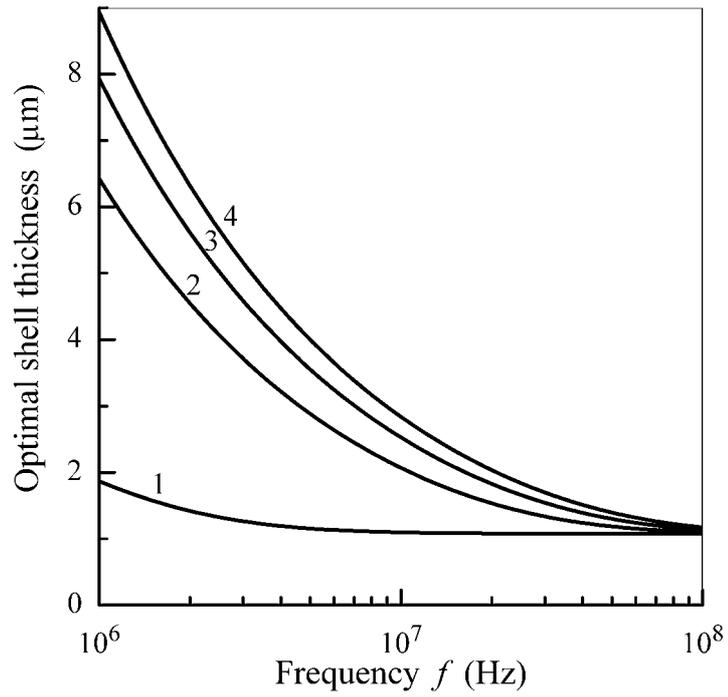

Fig. 7. Optimal shell thickness as a function of the current frequency at different anisotropy axis angle $\psi$: (1) $\psi=0$; (2) $\psi=0.05\pi$; (3) $\psi=0.1\pi$; (4) $\psi=0.15\pi$. Parameters used for calculations are $M=600$ Gs, $H_a=2$ Oe, $\sigma_1=5\times10^{17}$ s$^{-1}$, $\sigma_2=10^{16}$ s$^{-1}$, $\kappa=0.1$.